\documentstyle[12pt]{article}
   \advance\oddsidemargin by -1in
     \advance\evensidemargin
      by -1in
        \marginparsep
	 0.4cm
	    \advance\topmargin by
	     -0.0in
	       \makeindex
\def\lsim{\mathrel{\rlap{\lower4pt\hbox{\hskip1pt$\sim$}}
\raise1pt\hbox{$<$}}}

		        \newcommand\beq{\begin{equation}}
			 \newcommand\noi{\noindent}
			  \newcommand\eeq{\end{equation}}
			   \newcommand\beqn{\begin{eqnarray}}
			    \newcommand\eeqn{\end{eqnarray}}
			     \newcommand{\doublespace} {
			      \renewcommand{\baselinestretch} {1.6}
			       \large\normalsize}

\begin{document}
\vspace*{3cm}

\centerline{\Large \bf Baryon Number Flow in High-Energy Collisions}

\vspace{1cm}
\begin{center}
{\large Gerald~T.~Garvey}\\[0.3cm]
{\sl Los Alamos National Laboratory,
Los Alamos, NM 87545, USA}\\[1cm]
{\large Boris~Z.~Kopeliovich\footnote{On leave from
Joint Institute for Nuclear Research, Dubna,
141980 Moscow Region, Russia}
and Bogdan Povh}\\[0.3cm]
{\sl Max-Planck Institut f\"ur Kernphysik, Postfach
103980, 69029 Heidelberg, Germany}\\
\end{center}

\vspace{1cm}

\begin{abstract}
It is not obvious which partons in the proton carry
its baryon number (BN). We present arguments that BN is associated
with a specific topology of gluonic fields, rather than with
the valence quarks.
The BN distribution is easily confused
with the difference between the quark and antiquark distributions.
We argue, however, that they have quite different $x$-dependences.
The distribution of BN asymmetry distribution is nearly constant
at small $x$ while $q(x)-\bar q(x) \propto \sqrt{x}$.
This constancy of BN produces energy independence of the $\bar pp$ 
annihilation cross section at high energies.
Recent measurement of the baryon asymmetry at small $x$ at
HERA confirms this expectation. The BN asymmetry at 
mid-rapidities in heavy ion collisions
is substantially enhanced by multiple interactions, as has been
observed  in recent experiments at the SPS. 
The same gluonic mechanism of BN stopping increases the production
rate for cascade hyperons in a good accord with data.
We expect nearly the
same BN stopping in higher energy collisions at RHIC and LHC.
\end{abstract}

\doublespace

\newpage

\section{Introduction}

It is not obvious what carries the baryon number (BN) in a proton.
It is clear that the BN of a hadronic system is given by the number of 
quarks minus the number of antiquarks divided by three, and that 
this number is conserved in a closed system. This definition,
naively applied readily leads to an association of BN
with valence quarks. Recall that the density of valence quarks in a 
baryon of flavor $i$, carrying a momentum fraction x is defined as 
$q^v_i(x)=q_i(x)-\bar q_i(x)$,
\beq
\sum\limits_i\,\int\limits_0^1dx\,
\Bigl[q_i(x) - \bar q_i(x)\Bigr]= 
\sum\limits_i\,\int\limits_0^1dx\,
q^v_i(x) = 3 \label{0}
\eeq
which motivates the misconception that the valence quarks carry the
the BN. This is certainly is not correct. If one considers the reaction
\beq
\pi^- + p \to \Omega^- +K^+ + 2\,K^0\ ,
\label{0a}
\eeq
it is clear that BN is conserved, but none of the valence quarks
 in the initial proton appear as valence quarks in the $\Omega^-$.
This example holds a clue. The gluon fields create three $\ s$ and three
$\bar s$ quarks. It seems that no baryon number has been produced but 
the $3\bar s$ couple with $2d$ and a $u$ to form mesons and
$3s$ quarks form the $\Omega^-$. Thus BN must be carried by some other 
partons in the proton, probably gluons.

\medskip

\begin{figure}[tbh]
\includegraphics{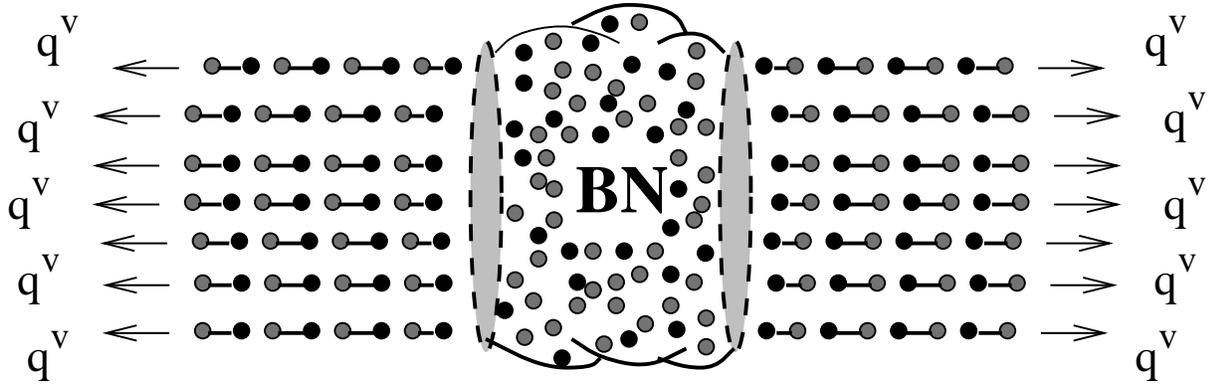}
\begin{center}
\vspace{5cm}
\parbox{13cm}
{\caption[Delta]
{\sl A cartoon illustrating the final state
following central collision of  (Lorentz contracted) relativistic
heavy ions. Grey and black circles correspond to 
quarks and antiquarks respectively.
The valence quarks $q^v$ escaping the collision region
produce jets consisted mainly of mesons. The considerable fraction
of BN of the projectile nuclei stops at mid rapidities.}
\label{fig:aa}}
\end{center}
\end{figure}

Another example which creates doubt that valence quarks 
carry BN is the
central collision of two heavy ions at very high energies
(RHIC, LHC). Naively, in such  high energy collisions one expects 
to find near zero net baryon number at mid-rapidities. This is because the valence
quarks of the colliding nuclei are difficult to stop, indeed energy loss
of a quark propagating through a heavy nucleus is known 
to be rather small, $\Delta E\sim
10\,GeV$, and energy independent 
\cite{cnn} - \cite{e866}. If the valence quarks are the  
carriers of baryon number they will sweep the baryon number along to large 
positive and negative rapidities.
We believe that the picture is true for the valence quarks but not for 
BN. The final state emerging from such a collision is 
shown schematically in Fig.~\ref{fig:aa}. 
A substantial fraction of the  
initial energy of the nuclei is stored in the valence quarks.
The rest of the energy is carried by many softer gluons and quark, anti-quark
pairs. A high-energy quark cannot be stopped by the soft interactions
it encounters in passing through a heavy nucleus. The valence quarks
pass through the collision region losing only a small
fraction of their  initial energy via gluon radiation induced by multiple
soft collisions. Many softer $\bar qq$ pairs and gluons are left behind 
as shown in Fig. 1. The energy density created by these
soft partons that are left behind is believed to be sufficiently
high, that a quark gluon plasma is created.

On the other hand, after propagation through a heavy nucleus
the initial valence quarks completely loose
their identity as a constituent of a nucleon. Therefore,
these quarks independently emerge from the collision and produce
fragmentation jets correlated with the colliding beam direction.
The jets will consist mostly of mesons and a small number 
of baryon-antibaryon pairs. Therefore the BN carried by the colliding 
nuclei is not to be found in the beam fragmentation regions but is stuck 
in the glue near mid-rapidity.

Assuming that the above illustrations are correct, a few conclusions 
immediately follow.\\
(i) The valence quarks readily survive this collision and remain
in the fragmentation regions of the projectile nuclei, while
the BN does not. Therefore valence quarks are not the carriers of BN.\\
(ii) BN stopping does not necessarily correspond to energy stopping. 
The energy carried by the valence quarks may penetrate through the nuclei
but the BN does not.
Of course if with a tiny probability 
the energy is completely stopped then the BN is also stopped.\\
(iii) Because the BN is stopped along with the glue, it appears that the 
gluon field may carry the BN. 

The latter conjecture has been around for quite some time. Some 25 years
ago the concept of a gluonic string junction as the carrier of BN 
was proposed \cite{artry,rv}.  Interest in this subject has waited 
over intervening two decades and has only intermittently been
discussed (for the latest review see \cite{rev})  
The early papers referred to above were written in the context of 
dual topological theories, Regge theory, and first generation QCD models. 
Current readers have little familiarity with these topics so 
these earlier presentations are difficult to follow or evaluate. The topic 
however has enjoyed renewed interest because of experimental results from HERA, 
which we will discuss later, and the recognition of its relevance to critical 
aspects of relativistic heavy ion collisions. In this latter instance, it 
is widely held that because the valence partons of the nucleons readily 
pass through the severely Lorentz contracted collision volume that they 
would emerge carrying most of the BN. This description would create a 
very hot region of strongly interacting matter containing small BN.
However, in the heavy ion collisions observed at the CERN SPS 
there are a far greater number of baryons at mid-rapidity than expected
assuming that BN is associated with valence quarks. 
Not only that, but a surprising number of these baryons contain strange 
valence quarks. In what follows we do not present new physics, but only 
attempt to make more accessible the idea that BN  is not tied to valence 
quarks, to show how BN  slides back to mid rapidity, and to show 
where experimental results bear out these ideas.

\section{Probes for the BN distribution}

To experimentally investigate the questions surrounding the nature of BN
one needs to identify suitable probes and reactions. Inclusive deep-inelastic 
scattering is clearly not useful as it probes only the distribution of 
electric charge and is insensitive to either gluons or BN.
Not surprisingly, the earliest probe used to investigate BN was the 
$\bar pp$ reaction where the annihilation cross was measured.
Another and more satisfactory way to study the x distribution of BN 
is to measure the BN asymmetry of produced particles. We will employ 
the usual assumption 
that there is a strong correlation in rapidity between the initial 
partons and the BN of the hadrons they produce. 

\subsection{BN annihilation does not vanish at high energies}

 Experiments on BN annihilation via $\bar pp \to mesons$ were carried out 
in the 1970s. These results have been under intensive discussion for many years. 
We summarize some of the important observations and conclusions.
If BN is associated with gluons, an immediate consequence is that 
BN should occur over the entire rapidity scale because of 
the fact that gluons are vector particles. No example or physical
reason leads us to believe that non-perturbative effects would 
alter this consequence. In section~\ref{flow} we will illustrate how the
process works. For the present, it is useful to recall the behavior of a 
couple of processes that are mediated by gluon exchange. For example, 
all hadronic total cross sections are observed to be nearly energy independent,
and further there is a persistence of 
large rapidity gaps observed in diffractive processes 
(Pomeron exchange) at the Tevatron and HERA. 
These are just 2 examples of the evidence for the persistence of gluonic 
interactions over all rapidities. 

  Therefore, if we associate BN with gluonic configurations, it is natural to 
expect that it will be rather uniformly distributed in rapidity and that
the $\bar pp$ annihilation cross section will not vanish at high energy. 
                                               
\begin{figure}[tbh]
\includegraphics{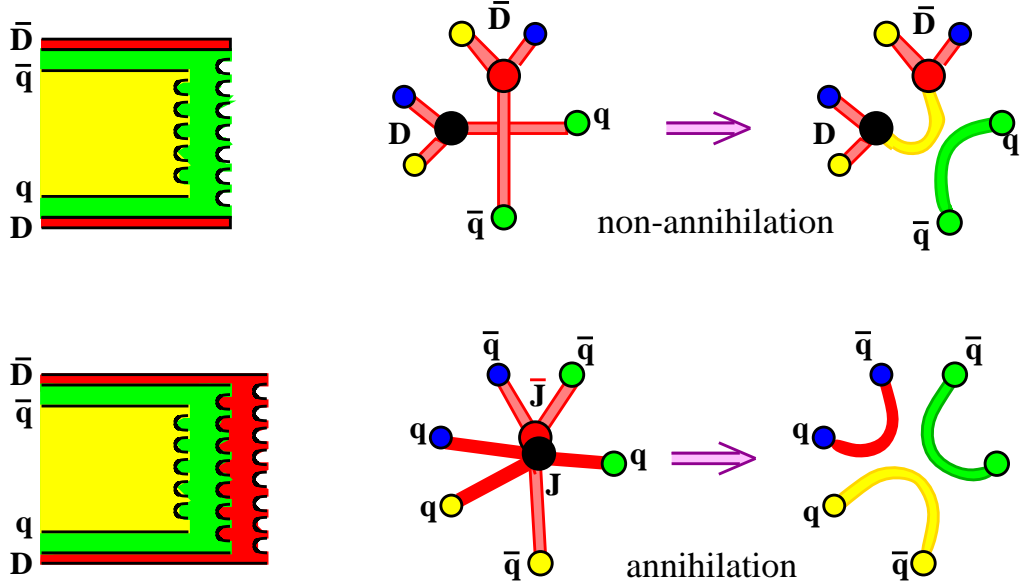}
\begin{center}
\vspace{7.7cm}
\parbox{13.5cm}
{\caption[Delta]
{\sl The cartoon shows interaction of a
proton consisted of a diquark ($D$) and a quark
with an antiproton. Crossing of the strings in
impact parameter plane leads to non-annihilation
final state with two strings (two sheet topology \cite{rv}).
Annihilation corresponds to overlap of $J$ and $\bar J$ leading
to three string production.}
\label{fig:nonpqcd}}
\end{center}
\end{figure}

The first claim that the $\bar pp$ annihilation cross section is
energy independent at high energies was made by
Gotsman and Nussinov \cite{gn}. They employed a string junction model
proposed earlier by Rossi and Veneziano, and suggested that annihilation results
from the overlap of a gluonic string junction and a 
string anti-junction followed by 
rearrangement of the gluonic strings as is illustrated in Fig.~\ref{fig:nonpqcd} 
on the bottom.  
They made a natural assumption that this process is energy
independent in analogy to nonannihilation collisions corresponding to
crossing of the strings as shown in Fig.~\ref{fig:nonpqcd} at the top. 
The asymptotic annihilation cross section was estimated by assuming 
that the string junction has a size of the order of the transverse 
dimension of the strings, $\sim
0.2 - 0.3\,fm$. With this assumption they found $\sigma^{\bar pp}_{ann}\approx 1 - 2\,mb$.

\begin{figure}[tbh]
\includegraphics{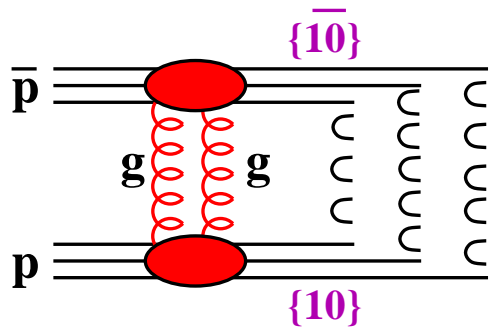}
\begin{center}
\vspace{4cm}
\parbox{11cm}
 {\caption[Delta]
{\sl Perturbative graph for $\bar pp$ interaction
via two gluon exchange in decuplet color state.
The three string final state leading to annihilation is
also shown.}
\label{fig:ann}}
\end{center}
\end{figure}

A confirmation of this pictorial description comes from a perturbative
QCD treatment of annihilation suggested in \cite {k,decameron}.
Annihilation in the $\bar pp$ interaction arises from multigluonic
exchange in a color-decuplet state as shown in Fig.~\ref{fig:ann}. 
As already mentioned, the cross section is energy independent 
since the gluon is a vector particle. 

The value of $\sigma^{\bar pp}_{ann}\approx 1 - 2\,mb$
predicted from a perturbative evaluation \cite{decameron}
of this annihilation cross section
$\sigma^{\bar pp}$  is the same as that obtained by Gotsman and Nussinov 
using the nonperturbative string junction approach.

Another, strong confirmation of this approach and the energy independence
of $\sigma^{\bar pp}_{ann}$ comes from the analysis of
the difference between the multiplicity distributions in $\bar pp$ and $pp$.
The difference in multiplicity is due to the specific three-string
topology of the events with string junction exchange \cite{rv}. They have a
multiplicity 1.5 times the non-annihilation events as one can see from
Fig.~\ref{fig:ann}. 
Using the enhanced multiplicity of the asymptotic purely gluonic annihilation
process, the asymptotic and asymptotic annihilation processes has been 
separated. The result of the analysis,
\cite{decameron} $\sigma^{\bar pp}_{ann}\approx 1.5\pm 0.1\,mb$ 
agrees very well with theoretical expectations. The data used in
the analysis included lab energies ranging from $10$ to $1480GeV$, 
and beautifully confirm the energy independence of this mechanism.
  Thus we have shown using theoretical analysis of experimental data
that the purely  asymptotic mechanism of BN transport over large 
rapidity intervals is rapidity independent. This means that the BN
distribution function at small x is proportional to 1/x, similar to sea 
quarks and gluons.

\subsection{BN asymmetry of produced particles}\label{asymmetry}

Another probe of the BN distribution is the BN asymmetry of 
produced particles which we define as
\beq
A_{BN}(x)=2\,\frac{N_{BN} - N_{\overline{BN}}}
{N_{BN}+N_{\overline{BN}}}\ .
\label{4}
\eeq
Here $N_{BN({\overline{BN}})}$ is the density 
of produced BN (anti-BN) which is a function of Bjorken $x$. 
We consider a case with initial $BN=1$ (proton - 
meson (photon) collisions). 

The BN asymmetry  (\ref{4}) can be interpreted as a ratio of production
rate of stopped BN and BN created from vacuum
(this is why we have factor 2 in (\ref{4})). This is correct
only if $A_{BN}(x)\ll 1$, {\it i.e.} if the effect of stopping
is relatively small.

It is natural to assume that the BN asymmetry of produced particles
arises from the BN asymmetry of the parton distribution in the
projectile proton. Then, the observation of an excess of BN produced at a
rapidity far below that of the projectile proton can be treated as a
measure of the BN at small x ($<10^{-3}$) in the partonic distribution 
of the proton.                                      

\begin{figure}[tbh]
\includegraphics{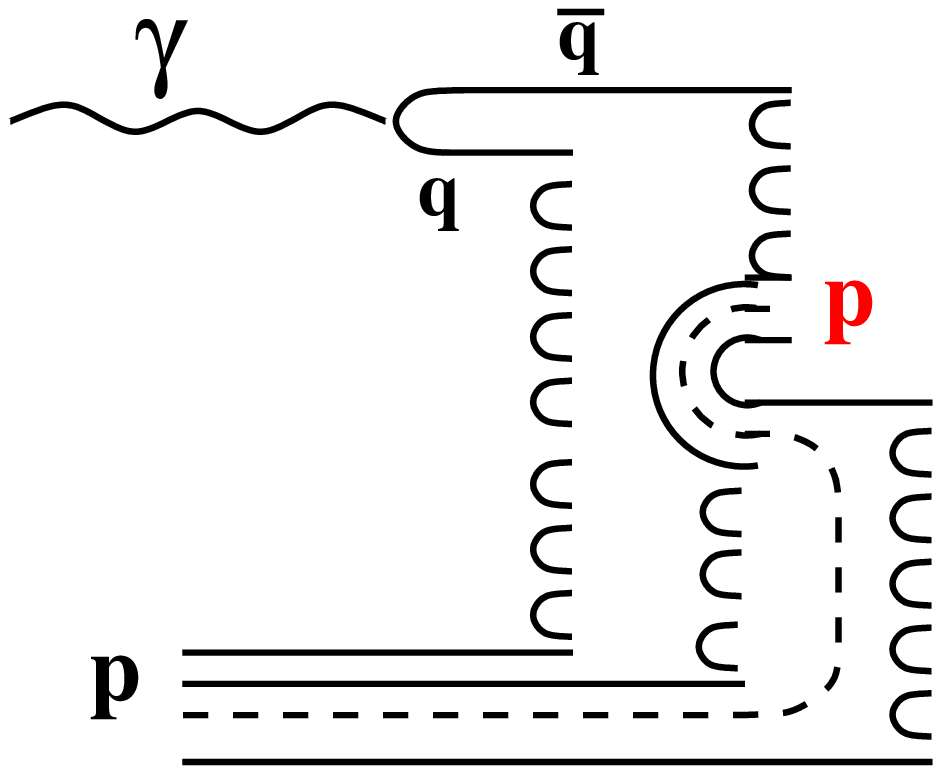}
\begin{center}
\vspace{5cm}
\parbox{13cm}
 {\caption[Delta]
{\sl The incident photon develops a BN-symmetric fluctuation
containing a baryon-antibaryon ($J\bar J$) pair. Annihilation of
the anti-BN of the fluctuation with the target BN leads to baryon asymmetry
in the photon fragmentation region.} 
\label{fig:gamma}}
\end{center}
\end{figure}

As the energy dependence of the annihilation cross section is known 
one can readily predict the expected BN asymmetry. Let us treat the 
case of a meson-nucleon collision viewed from the rest frame 
of the nucleon. The incident high energy meson develops a parton cloud containing  
fluctuations of baryon-antibaryon pairs ($J-\bar J$) with low probability.
Of course, this parton cloud is locally BN symmetric.
However, the antibaryon fluctuation in the meson can annihilate 
with the BN of the target nucleon, as is illustrated in Fig.~\ref{fig:gamma},
so that the partonic
fluctuation of the meson now has BN. The resulting BN asymmetry is given by
\cite{kp1},
\beq
A_{BN}(x)= \frac{\sigma_{ann}(s=m_N^2/x)}
{\sigma^{MN}_{in}}\ ,
\label{5}
\eeq
where $\sigma^{MN}_{in}$ is the inelastic meson-nucleon cross section.
Using the asymptotic value of $\sigma^{\bar pp}_{ann}=1.5\,mb$, and
$\sigma^{MN}_{in}=20\,mb$, a BN asymmetry, $A_{BN}=7\%$ was predicted 
in ref. \cite{kp1} for large rapidity intervals. 

Of course, a partonic treatment of the space-time description of an 
interaction is not Lorentz invariant, but depends on the reference frame.
Only observables are invariant under Lorentz transformations.
The same process of BN production in a $MN$ collision looks different
in the rest frame of the meson. In this case the nucleon develops  
parton fluctuations which are BN asymmetric down to the smallest $x$. 
This leads to a BN asymmetry of produced particles, which is also given by 
Eq.~(\ref{5}).

\begin{figure}[tbh]
\includegraphics{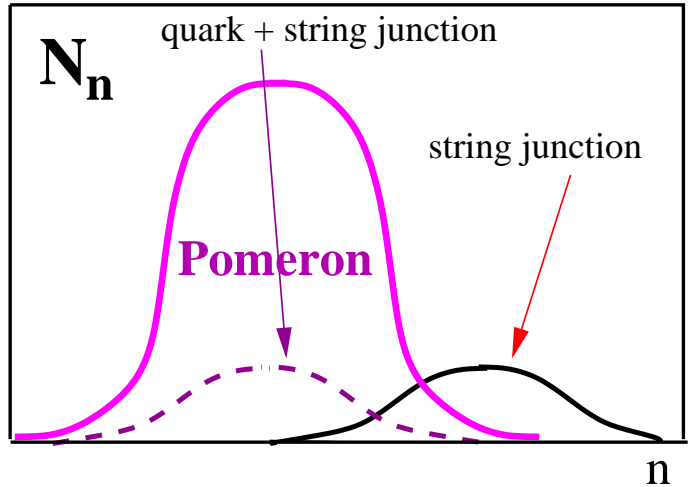}
\begin{center}
\vspace{7cm}
\parbox{13cm}
 {\caption[Delta]
{\sl Multiplicity distribution for produced pions corresponding to
the baryon symmetric contribution (Pomeron) dominating
the denominator in (\ref{4}) and baryon asymmetric mechanisms
related to quark-junction and single junction transfer
contributing into the numerator in (\ref{4}).}
\label{fig:multiplicity}}
 \end{center}
 \end{figure}

\begin{figure}[tbh]
\includegraphics{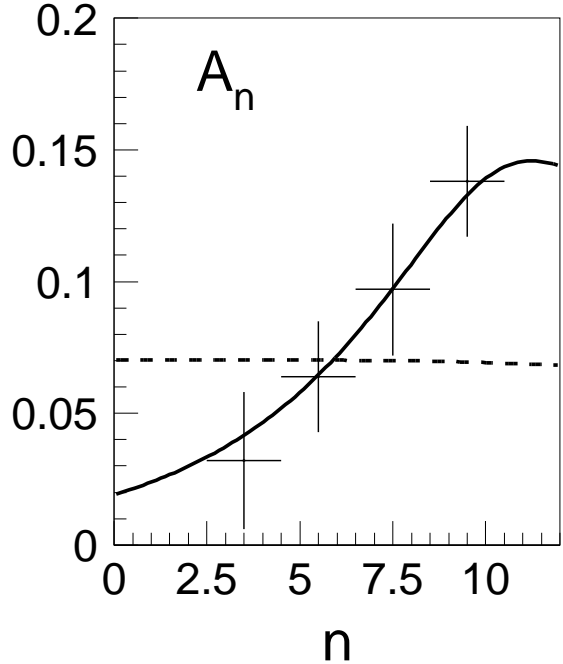}
\begin{center}
\vspace{8.6cm}
\parbox{13cm}
{\caption[Delta]
{\sl Dependence of baryon asymmetry (\ref{4}) on multiplicity
of produced particles. Solid and dashed curves show results
of calculations of \cite{kp2} assuming pure gluonic or
preasymptotic quark exchange mechanisms respectively. 
Data are from \cite{h1}.}
  \label{fig:h1}}
   \end{center}
    \end{figure}

Recently the H1 Collaboration at HERA \cite{h1} performed a measurement
of the BN asymmetry using the $\gamma-p$ interaction. The BN asymmetry is
observed in the photon hemisphere so that the rapidity interval from the
proton beam is large($\sim 8$ rapidity units).  The preliminary result 
$A_{BN}(x=10^{-3})=8\pm1\pm2.5\,\%$ in a good agreement
with the predicted value. Although the contribution to $A_{BN}$
of the preasymptotic mechanism (quark plus gluonic junction)
\cite{kz-zpc} vanishes as $\sqrt{x}$,
it still might be important at $x=10^{-3}$ \cite{kp1}. Once again, the two
mechanisms can be distinguished by the dependence of $A_{BN}$
on the multiplicity $n$ of produced particles. Indeed, the stopping of a string
junction requires a three string configuration which
produces a higher multiplicity of final particles than
a two string configuration, corresponding to the case when
the string junction is accompanied by a valence quark. 
The baryon symmetric contribution dominating the denominator in
baryon asymmetry definition Eq.~(\ref{4}) is related via unitarity to the
Pomeron which also corresponds to two-string in the final state. 
This difference between the two mechanisms in multiplicity distribution 
is illustrated in Fig.~\ref{fig:multiplicity}. 
As a result, the relative contribution of the asymptotic, gluonic mechanism  
of BN transfer increases as function of multiplicity. The results of
the analysis performed in \cite{kp2} for the dependence of baryon asymmetry
on the multiplicity of produced particles are compared with 
data from the H1 experiment \cite{h1} in Fig.~\ref{fig:h1}.
Apparently, the preasymptotic quark-exchange mechanism is excluded, while
the gluonic mechanism is in a good accord to the data.

This result from the H1 experiment at HERA is the first experimental 
evidence for a large BN asymmetry in the proton sea at small $x$.
Similar measurements can be performed with virtual
photons, but we expect no significant dependence on $Q^2$.

\subsection{Hyperon production}\label{hyperons}

It is also possible to measure BN stopping
via production of hyperons which are sometimes easier identified. 
It may help to discriminate between the two
mechanisms, asymptotic and preasymptotic.
If BN flows via the gluonic mechanism the string junction
picks up only sea quarks to create the 
final baryon. This leads to 
enhancement of hyperon production with relative 
statistical weights
$8/27$, $4/9$, $2/9$ and $1/27$ for nonstrange baryons, 
$\Lambda$ or $\Sigma$, $\Xi$ and $\Omega$ respectively.
It would be different, $4/9$, $4/9$, $1/9$ and $0$, respectively,
if the mechanism is preasymptotic one, associated with one of the 
valence quarks. On top of that there is of course a dynamical
suppression factor $w^n$, where $n$ is 
the number of strange quarks in the baryon,
and $w\approx 0.3$ is relative production probability of 
a strange quark from vacuum. 

The first measurements by the H1
Collaboration \cite{h1-lam} of BN asymmetry
for $\Lambda$ hyperons resulted in $A_{\Lambda}<0$
compatible with zero. Precise measurements of baryon asymmetry
for hyperons ($\Lambda^0,\ \Xi,\ \Omega,\ \Lambda_c$) were performed 
recently by the E791 Collaboration in collisions 
of $500\,GeV$ negative pions with carbon nuclei \cite{e791}. 
There are a few comments concerning these data which follow
in order.
\begin{itemize}
\item
The c.m. energy in this experiment is twice as low as the maximal energy 
reached in $pp$ collisions at ISR in the experiments studying the 
$p/\bar p$ asymmetry at mid rapidity \cite{isr1}
and rapidity dependence of this asymmetry \cite{isr2}.
According to calculations in \cite{kz-zpc}
the observed asymmetry in both experiments is dominated by the preasymptotic
mechanism of valence quark and string junction exchange, and this is 
also true for the E791 data.
\item
The definition of asymmetry used in \cite{e791} is different
from our (\ref{4}), it does not include factor 2.
Therefore all measured asymmetries should be doubled to compare
with our predictions and with the data from ref.~\cite{h1}.
\item
The nuclear target can modify BN asymmetry at negative $x_F$.
\item
The data demonstrate a rising dependence of asymmetry on 
transverse momentum. This is not surprising since both
preasymtotic and asymptotic mechanism involve breaking of 
the projectile diquark in the nucleon. There are many 
experimental and theoretical evidences \cite{diquark} that
the nucleon wave function is dominated by a component with a 
compact diquark with separation $\sim 0.3\,fm$. In this case
diquark destruction leads to liberation of rather large
intrinsic transverse momenta of the quarks.

\end{itemize}

According to the partonic interpretation in Section~\ref{asymmetry}
illustrated in Fig.~\ref{fig:gamma} the probability of BN flow
(the numerator in (\ref{4})) 
is proportional to the number of $BN,\ \overline{BN}$ pairs created
from vacuum. The latter is the denominator in (\ref{4})
provided that $A_{BN}\ll 1$. In this case the number of
sea $BN,\ \overline{BN}$ pairs cancels and the BN asymmetry can be
treated as a measure of BN stopping.

At this point we should distinguish between the BN asymmetry 
$A_{BN}(x)$ and the asymmetry 
\beq
A_{B}(x)=2\,\frac{N_B-N_{\bar B}}{N_B+N_{\bar B}}
\label{5a}
\eeq
for a partiqular species of baryon, $B$.
The produced BN (string junction) is realized via production of
a variety of baryons with corresponding relative branchings. 
Since the mechanisms of hyperon creation from vacuum (the denominator in
$A_B(x)$) and via the stopped string junction (the numerator in $A_B(x)$)
may be quite different, the baryon asymmetries for hyperons (B=H)
will differ from the the BN asymmetry (\ref{4}). For instance,
if the contribution of BN flow to production of
a hyperon is substantially greater  the one from
vacuum, a specific baryon asymmetry (\ref{5a}) will
approaches the maximal value 2, while the BN asymmetry (\ref{4})
may be small. This may happen with multi-strange hyperons
if the gluonic mechanism dominates. Indeed, production
of multi-strange
hyperons from vacuum should be greatly suppressed by 
the string mechanism for BN production suggested
in \cite{cnn}.
In this case $A_B(x)$ is not a characteristic of BN
flow.

Calculation of baryon asymmetry for hyperons is more difficult problem 
than for BN asymmetry (\ref{4}) especially at medium high
energies. It requires decomposition into specific baryons of the
BN produced from vacuum.
A better characterization of BN flow would be ratios of
$N_B-N_{\bar B}$ for different baryon species  
normalized by the same yield of vacuum pairs, {\it e.g.}  
$N_p+N_{\bar p}$.

\section{How BN Flows}\label{flow}

We shall now illustrate how BN can range over a very large interval in
x, and in particular how it can appear at very small x.

The string configurations of the color fields in a meson 
and in a baryon are quite different. A meson looks like a $\bar qq$ 
pair connected by a color flux tube \cite{cnn}. The 
configuration of strings in a baryon having minimal energy
has a form of the Mercedes-Benz star (see Fig.~\ref{fig:nonpqcd}), 
and the point where the strings join 
is called string junction (J) \cite{artry}. This configuration 
of gluonic fields also follows from the gauge invariant form 
of the operator with $BN=1$ as suggested in \cite{rv}.
Correspondingly, an anti-string-junction may also be introduced
($\bar J$) as the source of antibaryons.
$J$ and $\bar J$ can interact and annihilate to mesons.

The association of BN with the topology of gluonic fields 
rather than with quarks is not new. 
A topological view of BN was suggested a long time ago in the 
chiral soliton model of Skyrme \cite{skyrme}. The Lagrangian of 
the so called Skyrme Model does not even contain quark degrees
of freedom.

An assignment of BN to a string junction is quite compatible
with the results presented in Section~2 which showed that
BN displayed characteristics normally associated with gluons. 
Of course a gluon does not have any BN, and the dynamical 
association of BN with gluonic fields is explained 
in following paragraphs.

\begin{figure}[tbh]
\includegraphics{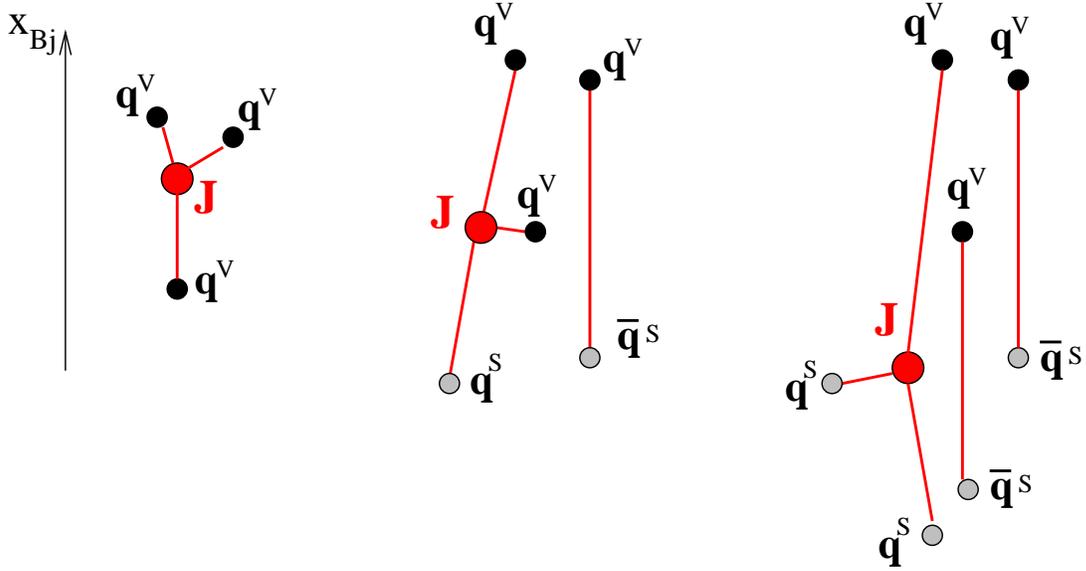}
\begin{center}
\vspace{7cm}
\parbox{13.5cm}
{\caption[Delta]
{\sl String configurations
corresponding to different terms
 in Fock state
 decomposition (\ref{1}).  Grey and black circles
 show the sea and 
 valence quarks respectively.  
 The dotted lines show the 
 color strings. Conventionally we assume that the
 vertical axis corresponds to Bjorken $x$.}
\label{fig:fock}}
\end{center}
\end{figure}

In the infinite momentum frame the string junction shares the proton
momentum, therefore, it can be given a  partonic interpretation.
Since a Fock state decomposition of the proton 
contains components with a few sea $\bar qq$ pairs,
\beq
 |p\rangle = |3q^v\rangle + |3q^vq^s\bar q^s\rangle +
|3q^v2q^s2\bar
 q^s\rangle + ...\ ,
\label{1}
\eeq
three sea quarks can form a baryon by extending the processes illustrated
in Fig.~\ref{fig:fock}. 
Thus it is conceivable that BN might have a distribution 
$\propto 1/x$ at   
small x similar to that of gluons and sea quarks \cite{kp1}.

\begin{figure}[tbh]
\includegraphics{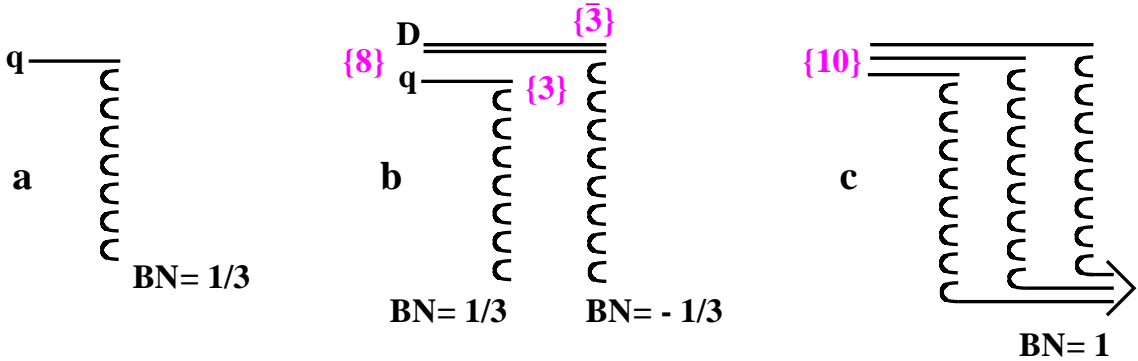}
\begin{center}
\vspace{4cm}
\parbox{13.7cm}
{\caption[Delta]
{\sl The $\bar qq$ chain ({\bf a}) effectively transports
the BN of the valence quark down to small $x$
(Bjorken $x$ is assumed to follow the vertical axis).
If, however, the valence quark and diquark form
a color octet in final state ({\bf b}), 
no BN is transferred to small $x$ region.
Nevertheless if the valence quarks are in color decuplet state
({\bf c}) the BN is transported by three $\bar qq$ chains.}
\label{fig:chains}}
\end{center}
\end{figure}

  Though at first glance it appears simple, it is not trivial 
to push the BN in a baryon like the proton down to small $x$. One 
can imagine the parton cloud of a valence quark as a chain of $\bar qq$ pairs
as shown in Fig.~\ref{fig:chains}a.
The terminus of this chain is, of course, a quark with BN=1/3.
In the proton this valence quark is accompanied by a valence diquark.
The valence diquark has a color wave function $\{\bar 3\}$, the same
as a $\bar q$ in order that the proton's valence wave function u(ud)
be an over all color singlet. The valence diquark therefore
develops a $\bar qq$ chain that terminates in a $\bar q$. Thus the net 
BN=0 at small $x$ as shown in Fig.~\ref{fig:chains}b. Only if the three valence 
quarks get into a decuplet color state can they propagate BN=1 down to
small x (Fig.~\ref{fig:chains}c). The decuplet state can be created in higher Fock 
components of the proton such as, say $\ uudg \bar qq$ where the color 
degrees of freedom of the $\bar qqg$ allow the uud valence configuration
to be a decuplet. In this instance the three $\bar qq$ chains can transport
BN=1 to small $x$. The probability to produce a color decuplet
$3q$ state in hadronic collisions turns out to be quite small,
as evaluated in ref.~\cite{decameron}.

A few remarks are in order. It is clear that the $\bar qq$
chain while carrying the baryon number of its origin does not have to carry
its flavor. Thus the baryon generated at low x need bear no resemblance
to the the valence baryon, for example $uud \to sss$ as we discussed earlier. It is quite
clear that in a real sense the $\bar qq$ chain carries the baryon
number of its origin. Thus the idea that the junction of three such chains
is the topological realization of BN seems quite reasonable and does not
contradict any known fact. Of course the mechanism discussed above is not
the only way in which BN moves to lower x. Returning to Fig.~\ref{fig:chains},
usually BN emerges from a collision carried by the valence diquark which
picks up another quark from the vacuum. Such a process would have the
resulting BN at relatively large x ($\ x>0.1$). On the other hand if two of the 
quarks are from the sea then the x of the BN is brought much 
lower.  This can be interpreted as a purely gluonic transfer,
of just the junction, and only then does it become independent of x. As this mechanism
is much smaller than the two previously mentioned, it can only be clearly 
observed at very low x ($\ x< 10^-3$) where the diquark- and preasymptotic
quark-exchange mechanisms have disappeared.

\section{Heavy ion collisions}\label{sect:nuclei}

The pedagogical example of central collisions of relativistic heavy ions
mentioned in the introduction is in fact more pertinent than simply as 
an illustration. It shows that at high energies the momentum of 
the projectile valence quarks survive the multiple interactions 
in the collision, but loose their identity as baryonic constituents. 
This means that BN of the projectile nuclei moves from the projectile 
fragmentation regions and accumulates at central rapidities.
This BN flow moderated by external gluons \cite{kz-zpc} should lead 
to a considerable baryon stopping at mid-rapidities. 

\begin{figure}[tbh]
\includegraphics{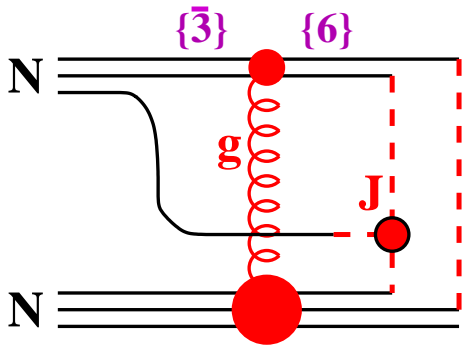}
\begin{center}
\vspace{5cm}
\parbox{13cm}
 {\caption[Delta]
{\sl Interaction of a fluctuation of the incident proton
  containing a valence quark and the string junction
  at small x (with a small probability $\propto \sqrt{x}$)
  Interaction is illustrated using a perturbative gluonic
  exchange for the sake of simplicity.}
  \label{fig:stoppinga}}
  \end{center}
  \end{figure}

The BN observed so far in proton-proton collisions at SPS and ISR 
remains primarily in the nucleon's fragmentation regions
and is only found at smaller rapidities because of the spread in
the momentum fractions of the valence quarks \cite{kz-zpc} as is sketched
in Fig.~\ref{fig:stoppinga}. 
A fluctuation of the incident nucleon
into a fast diquark and relatively slow quark interacts with the
target (gluon exchange is used for illustration). As a result the longitudinal
momenta of the valence quarks remain unchanged, but the diquark may
switch from anti-triplet to sextet color state. In this case the projectile
BN acquires the rapidity of the slow valence quark. The cross section dependence
on the rapidity interval $\propto\exp(-\Delta y/2)$ follows the primary
momentum distribution of the valence quark. The perturbative estimate
of the absolute value of the distribution performed in 
\cite{kz-zpc} is in a good agreement with data.
The same mechanism of diquark breaking applied to $\bar pp$ annihilation
\cite{kz1} also explains the data available up to energy $E\sim 10\,GeV$.

This input was
used in \cite{ck} to predict that almost all of the BN is stopped 
in lead-lead collisions at SPS. Why is the 
stopping so different in the heavy ion case? The probability for the
projectile color-anti-triplet diquark to survive multiple interactions
vanishes in a heavy nucleus, therefore the BN escapes from the
projectile fragmentation region with a probability close to one.
The shift of BN in such heavy ion collisions involves 
processes really very  different from the ones considered in Section~2 as 
subsequent collisions of the diquark in the nuclear medium are required to 
reduce the rapidity of the BN. Measurements by the NA49 collaboration 
\cite{na49} confirmed these expectations. 

Similar results were recently obtained \cite{miklos} 
using a Monte Carlo code that implements the notion of string junctions.
In spite of the claim in \cite{miklos,dima} that the 
asymptotic pure string-junction mechanism is used, it is only important  
that they assume a $1/\sqrt{s}$ energy dependence which is the same
as ref.~\cite{ck}.
We believe that prescription of value $\alpha^J_0(0)\approx 1/2$
to the Reggeon intercept corresponding to string junction exchange \cite{rv}
has no justification beyond citation of the result of \cite{eh}.
In this respect it worth quoting the authors Eylon and Harari:
``{\sl the crude model described here, is not meant to be taken seriously as 
a \underline{quantitative} description of $\bar BB$ processes.}'' \cite{eh}.
Another assumption of \cite{rv} that the asymptotic mechanism of string junction
exchange already dominates at energies $E<10\,GeV$ and 
identification of annihilation with the difference between the total $\bar pp$ 
and $pp$ cross sections, contradicts the basic ideas of ref.~\cite{eh} and
was criticized in \cite{rev}. Disappearance of the $\omega$ exchange in $pp$ 
scattering would lead to serious troubles for the high-energy phenomenology.

\begin{figure}[tbh]
\includegraphics{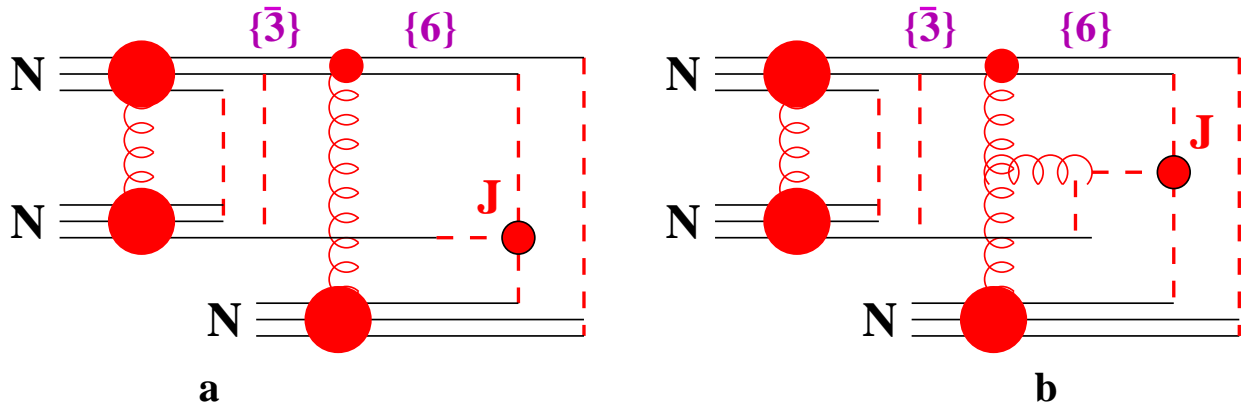}
\begin{center}
\vspace{5cm}
\parbox{13cm}
 {\caption[Delta]
{\sl Double interaction of a nucleon in the target nucleus.
The first interaction
breaks down coherence between the projectile quark and diquark.
The second interaction liberates the projectile BN converting the diquark
from the $\{\bar 3\}$ state to $\{6\}$. The BN acquires the rapidity of
a valence quark in the target nucleon ({\bf a}) or a radiated gluon ({\bf b}).}
\label{fig:stoppingb}}
\end{center}
\end{figure}

It is assumed in \cite{ck} that BN liberated via multiple interactions in
heavy ion collisions move to the rapidities of valence quarks
similar to $NN$ collisions as is shown in Fig.~\ref{fig:stoppinga}.
If it were true the probability to stop BN at central rapidities would
decrease with colliding energy as $s^{-1/4}$, {\it i.e.} one should expect
three times less stopping at RHIC compared to SPS. This is not obvious,
however. Indeed, the diquark and quark in the projectile nucleon
lose coherence in the very first inelastic interaction on the surface of
the nucleus and ``do not talk to each other'' any longer. The projectile BN
associated with the diquark and liberated in subsequent multiple interactions
as is shown in Fig.~\ref{fig:stoppingb}a should acquire the rapidity of a
valence quark of the target ({\bf a}) or a gluon (or a sea $\bar qq$)
radiated at mid rapidities ({\bf b}). 
Although the total probability of BN flow to mid rapidities is
energy independent, its sharing between the two mechanisms in
Fig.~\ref{fig:stoppingb}a,b depends on energy. 
At high energies it is more probable that the BN will be stuck
with one of the numerously radiated gluons (Fig.~\ref{fig:stoppingb}a),
while at medium high energies contribution of valence quarks 
(Fig.~\ref{fig:stoppingb}b) may be important. 

The energy independent mechanism illustrated in Fig.~\ref{fig:stoppingb}
was also discussed in \cite{ck},
although, it was assumed that it will take over only at very high 
energies, while the preasymptotic mechanism depicted in
Fig.~\ref{fig:stoppinga} still dominates at the SPS energies. 
However, it follows from the above consideration 
that on heavy nuclei double (multiple) step mechanism Fig.~\ref{fig:stoppingb}
should dominate at any energy.
Thus, we expect nearly a full baryon stopping in central gold-gold 
collisions at RHIC as it was observed in lead-lead collisions at SPS.
The stopped BN should be spread over the whole rapidity range.

A sensitive test for the BN stopping mechanisms is the fraction of hyperons
produced at mid rapidities, especially cascades, as is
discussed in Section~\ref{hyperons}. Suppression of double-strange hyperons
relative to non-strange baryons,
$R(\Xi/N)=(N_{\Xi}-N_{\bar\Xi})/(N_N-N_{\bar N})$, was measured  in the NA49
experiment for central Pb-Pb collisions 
at SPS CERN \cite{na49,na49-xi} at $R(\Xi/N)=0.063 \pm 0.006$,
what is much higher that in $pp$ and $pA$ collisions.
According to statistical and strangeness
suppression factors presented in Section~\ref{hyperons}
we expect for the pure gluonic mechanism depicted in Fig.~\ref{fig:stoppingb}b
$R(\Xi/N)\approx 0.067$ in a very good agreement with the data.
At the same time both mechanisms shown in Figs.~\ref{fig:stoppinga}
and \ref{fig:stoppingb}a which associate BN stopping with
valence quarks predict three times smaller value $R(\Xi/N)\approx 0.022$.
In order to explain the experimental value it was assumed in \cite{capella}
that additional $\Xi$s are produced via final state cascading, 
$\Lambda+\pi\to\Xi+K$, which is difficult to evaluate.

BN stopping in $\Omega$-hyperon channel is even more sensitive
to the mechanisms under discussion. Unfortunately, no data are
available yet, only the
sum $N_{\Omega} + N_{\bar\Omega}$ was measured in \cite{na49-omega}.

One should be cautious  interpreting enhancement of hyperons in heavy ion
collisions as an indication to a new physics. The conventional mechanisms
of BN stopping well explain the data.

A possibility of weak energy dependence for baryon 
stopping in heavy ion collisions has been also discussed 
recently in \cite{bopp} basing on the topological treatment
for baryon flow. In this approach enhanced BN stopping arises from
fusion of abundantly produced Pomeron cylinders into so called
membraned cylinders. Although this scheme, similar to \cite{rv}
suggests a nice geometrical interpretation for BN flow, it is
not suitable for numerical calculations and does not lead to any
concrete predictions.

It is worth mentioning that
there is still an exotic possibility that three valence quarks
of the projectile nucleon will retain the BN in the final
state with a substantial probability.
Indeed, the three quarks after they have traveled through a heavy nucleus 
and experienced multiple interactions are completely unpolarized in color
space, {\it i.e.} they are in a color decuplet, octet or singlet state with 
probabilities $10/27$, $16/27$ and $1/27$ respectively.
Thus, with probability $17/27$ they may retain a BN. This does not contradict 
the previous consideration of multi-step interaction which is supposed to
move BN from the projectile to mid rapidities. In the same way
multiple interactions can move BN back to the projectile leaving at mid
rapidity a baryon-antibaryon pair. In this scenario only a fraction $10/27$
of the BN stored in the colliding nuclei can stop, and extra
baryon-antibaryons are produced. 
Nevertheless, it seems to be very unrealistic to believe that the three
quarks which are hadronizing independently can come together and 
create a color octet or singlet state. We assume that it never happens,
although it should be tested in $pA$ collisions at high energies.

\section{Summary, conclusions and outlook}

In this note we have rejected the widely spread view
that BN should be associated with the valence quarks of a 
baryon. We also show that BN cannot be identified with
the difference $q(x)-\bar q(x)$, because BN has a
different small $x$-dependence. That is, the $q-\bar q$ difference
vanishes at small $x$ while the BN asymmetry does not.
The latter statement is supported by the available data and theoretical
arguments of both perturbative and nonperturbative origin,
the most convincing being the net gluonic mechanism for BN transfer
which contributes about $1.5\,mb$ to the difference in the  multiplicity
distributions in $\bar pp$ and $pp$ collisions. Such a rapidity
independent cross section is in excellent agreement with the $8\%$
BN asymmetry at $x<10^{-3}$ and its multiplicity
dependence recently measured at HERA.

Consideration of baryon structure in a string model and
mechanisms of BN transfer to small $x$ leads to the notion that BN
is associated with a specific topological configuration of
gluonic fields rather than with quarks. Such a concept is consistent
with a similar view underlying the chiral soliton description of 
baryons \cite{skyrme}. 

A natural question to address, is what further experimental studies 
should be done to test, clarify and extend the ideas presented here.

\begin{itemize}
\item 
It is very important to verify the $x$-independence of the BN
asymmetry at small $x$ by extending the measurement done at HERA 
\cite{h1}. Closely related experiments can be also be carried out
at the Tevatron.
It is also important to measure baryon asymmetry 
for hyperon production in such experiments, 
since it is sensitive to the mechanisms
under discussion. 
\item
As the gluonic mechanisms generating BN are flavor independent
one should see an excess of BN in every baryonic channel. Also 
this mechanism produces a much larger yield of strange and multi-
strange baryons in relativistic heavy ion collisions
than comes from conventional quark-diquark string models.
\item
A new mechanism for $\bar JJ$ string junction pair production
via color-decuplet gluonic (or string-junction)
exchange is suggested in \cite{kz-zpc}. It naturally 
leads to long range
rapidity correlations between produced baryons and antibaryons.
This should be tested at the Tevatron collider.
\item
 We expect the gluonic multi-step mechanism of BN stopping to
dominate in the SPS-RHIC-LHC energy range for the central collisions
of very heavy ions. In this case the rate of BN stopping,
$d(BN-\overline{BN})/dy$, decreases logarithmically with energy due
to the growth of the total rapidity interval. Although a reliable
prediction needs detailed calculations, one can estimate the energy
dependence of BN stopping as follows. Since the stopped string
junction can stick with any of radiated gluons, it will have the same
rapidity distribution as the gluons. It is known that gluons dominate
at Bjorken $x \lsim 0.1$. Therefore the BN stopped by the multi-step
gluonic mechanism will be distributed over a rapidity interval which
is about four units (two from each side) shorter than the whole
rapidity range accessible in the collision. It comes to about 2, 8
and 13 units of rapidity intervals at SPS, RHIC and LHC respectively.
Correspondingly, one should expect that the amount of net BN stopped
at RHIC and LHC compared to SPS to be suppressed by factors $0.25$
and $0.15$ respectively. In the case of the preasymptotic mechanism
associated with valence quarks \cite{ck,miklos} the expected relative
suppression of stopping is expected to be nearly the same at RHIC but
much smaller, $\sim 0.06$ at LHC. The expected shapes of rapidity
distributions are quite different. The BN stopping mechanism related
to the valence quark distribution leads to a distribution
proportional to $\exp(y/2)+\exp(-y/2)$ which has a minimum at the mid
rapidity $y=0$.
  \item The mechanism of string junction flow for BN stopping leads
to enhanced hyperon production, more for the asymptotic than
preasymptotic mechanisms. In heavy ion collisions the multi-step
mechanism of BN flow leads to the production rate of cascades ($\Xi -
\bar\Xi$) which agrees well with experimental data and is three times
higher than predicted by the stopping mechanism associated with
valence quarks. \item Since both mechanisms of BN sopping involve
destruction of the projectile diquark, the transverse momenta of
produced particles should be increased in such events. Both effects
are enhanced in central collision of heavy ions.

\end{itemize}. 

\noi
{\bf Acknowledgments:} we are grateful to Ulrich Heinz who has read the
manuscript and made many improving suggestions. We also thank Andrei Rostovtsev 
informing us about data \cite{e791}, and Johanna Stachel who has drawn our
attention to the data on production of cascade hyperons in heavy ion
collisions.

\end{document}